\documentclass[acmsmall]{acmart}

\AtBeginDocument{%
  \providecommand\BibTeX{{%
    \normalfont B\kern-0.5em{\scshape i\kern-0.25em b}\kern-0.8em\TeX}}}

\setcopyright{acmcopyright}
\copyrightyear{2023}
\acmYear{2023}
\acmDOI{XXXXXXX.XXXXXXX}

\acmJournal{TIIS}
\acmVolume{1} 
\acmNumber{1} 
\acmArticle{1} 
\acmMonth{1}
\acmPrice{15.00}
\acmDOI{10.1145/3484509}

\usepackage{xcolor, amsmath}
\usepackage{enumitem}
\usepackage{dsfont}
\usepackage{booktabs}
\usepackage{makecell}
\usepackage{multirow}
\usepackage{graphicx}

\begin{document}

\title{Break Out of a Pigeonhole: A Unified Framework for Examining Miscalibration, Bias, and Stereotype in Recommender Systems}

\author{Yongsu Ahn}
\affiliation{%
  \institution{University of Pittsburgh}
  \city{Pittsburgh}
  \country{United States}}
\email{yongsu.ahn@pitt.edu}
\author{Yu-Ru Lin}
\affiliation{%
  \institution{University of Pittsburgh}
  \city{United States}
  \country{United States}}
  \email{yurulin@pitt.edu}

\renewcommand{\shortauthors}{Ahn and Lin}

\begin{abstract}
  Despite the benefits of personalizing items and information tailored to users' needs, it has been found that recommender systems tend to introduce biases that favor popular items or certain categories of items, and dominant user groups. In this study, we aim to characterize the systematic errors of a recommendation system and how they manifest in various accountability issues, such as stereotypes, biases, and miscalibration. We propose a unified framework that distinguishes the sources of prediction errors into a set of key measures that quantify the various types of system-induced effects, both at the individual and collective levels. Based on our measuring framework, we examine the most widely adopted algorithms in the context of movie recommendation. Our research reveals three important findings: (1) Differences between algorithms: recommendations generated by simpler algorithms tend to be more stereotypical but less biased than those generated by more complex algorithms. (2) Disparate impact on groups and individuals: system-induced biases and stereotypes have a disproportionate effect on atypical users and minority groups (e.g., women and older users). (3) Mitigation opportunity: using structural equation modeling, we identify the interactions between user characteristics (typicality and diversity), system-induced effects, and miscalibration. We further investigate the possibility of mitigating system-induced effects by oversampling underrepresented groups and individuals, which was found to be effective in reducing stereotypes and improving recommendation quality. Our research is the first systematic examination of not only system-induced effects and miscalibration but also the stereotyping issue in recommender systems.

\end{abstract}



\received{20 February 2007}
\received[revised]{12 March 2009}
\received[accepted]{5 June 2009}

\maketitle

\newcommand{\yrl}[1]{{\color{red}{[YRL:#1]}}}
\newcommand{\ysc}[1]{{\color{blue}{[YS:#1]}}}

\newcommand{\ys}{\textcolor{blue}}
\newcommand{\ysv}{\textcolor{orange}}
\newcommand{\yrv}[1]{{\color{purple}{#1}}}
\section{Introduction}

Recommender systems have been widely adopted in a variety of domains, ranging from online shopping to high-stakes decisions, such as insurance or job applications. Using data and recent advances in AI techniques, these systems can generate ``personalized'' items and information that are tailored to the needs of individual users. Despite its benefits in reducing their browsing and retrieval efforts, studies have found that recommender systems induce biases, such as popularity bias \cite{ControllingPopularityBiasLearningtoRank}, which favors \textit{popular} items, filter bubbles \cite{Exploringfilterbubbleeffectusing} or echo chambers \cite{UnderstandingEchoChambersEcommerceRecommender}, where recommended items or users in the system are tailored by a group of users of \textit{similar} kind, resulting in item/group segregation. Recommender systems are also likely to amplify user preference for certain item categories (e.g., action or sci-fi movies) that are \textit{dominantly} preferred by a larger group of users (e.g., men) \cite{BiasDisparityRecommendationSystems, Crankvolumepreferencebiasamplification}. 

Compared with the issue of system-induced bias, the existence of stereotypes perpetuated by recommendation systems is relatively unexplored. ``Stereotype'' refers to the behavior of overgeneralizing user characteristics by regarding individual users as having typical preferences of a predetermined group (e.g., based on demographic or other attributes) \cite{StereotypesStereotypingMoralAnalysis}. For example, in movie recommendations, a system may recommend to male users a set of items that typical male users will like. Such behavior can lead to discrimination against minority subgroups or individuals with particular interests, restricting users' choices into typical preferences, which in turn reduces user satisfaction or the quality of recommendations. 

In this study, we look at the stereotypes and biases that arise as a result of data underrepresentation within a system. Because underrepresented users -- those with atypical preferences or who belong to minority groups -- are more likely to be impacted, they are more likely to receive poor-quality recommendations, resulting in discrimination against underrepresented subgroups or individuals. These system-induced behaviors can worsen the miscalibration problem, also known as the mismatch between user preferences and specific item categories  \cite{Calibratedrecommendations}.

While there is collective evidence pertaining to user characteristics \cite{ConnectionPopularityBiasCalibrationFairness, CalibrationCollaborativeFilteringRecommenderSystems}, systematic behaviors \cite{ControllingPopularityBiasLearningtoRank, ExposureIdeologicallyDiverseNews, Howalgorithmicconfoundingrecommendationsystems, sunstein2009going}, and miscalibration \cite{Calibratedrecommendations}, we find that existing literature lacks a comprehensive understanding of the underlying relationships, posing the following challenges: 1) A lack of a systematic approach to identifying the relationship between users, algorithms, and errors: Existing research has contributed to the identification of various types of system-induced biases; however, there is a lack of systematic methods to examine how these system-induced effects are associated with user characteristics and exacerbate recommendation quality. 2) Insufficient comprehension of the stereotyping issue in recommender systems: Unlike other potential system-induced effects, the problem of stereotyping has not been studied in the literature. Stereotyping recommendations can limit the diversity of individual preferences, especially by influencing users with atypical preferences to receive recommendation items that are typically preferred by the groups to which they belong, thereby failing to calibrate the user preferences.

This study aims to provide a unified framework for quantifying stereotypes, bias, and miscalibration, given the challenges and the need for a better understanding of the relationship between them. Using the proposed framework, we are able to analyze miscalibration at a granular level, examine the interconnectedness of system-induced effects, and gain a deeper insight into those systematic effects on both user groups and individuals. We focus on answering the following research questions: 1) \textbf{Sources of miscalibration}: To what extent do two distinct sources of error, bias, and variance, contribute to miscalibration? 2) \textbf{System-induced effects}: How does it relate to system-induced effect? and 3) \textbf{Disparate impact}: How does it affect groups and individuals disproportionally? Based on the proposed measures, we conduct an observational study on existing recommender systems to observe how systematic effects and stereotyping problems result in negative impacts on groups and individual users. Moreover, we investigate potential mitigation strategies for users negatively affected by the system-induced behavior. The contributions of this study are as follows:

\begin{itemize}
    \item \textbf{A unified framework for understanding miscalibration, bias, and stereotype}: We present a statistical framework for decomposing miscalibration into its two distinct sources, bias and variance, as well as their relationship with stereotype. This enables us to investigate further systematic behaviors that lead to various types of effects and have a disproportionate impact on underrepresented groups and individuals.
    \item \textbf{A formulation of the stereotyping issue in recommender systems}: We discuss the problem of stereotyping in recommender systems,  which is the tendency to overgeneralize user preferences based on their group membership. This behavior is associated with two system-induced effects, stereotype and inflated diversity, both of which contribute to miscalibration. This is the first study to examine the risk of stereotyping problem in recommender systems.
    \item \textbf{An analysis of the relationship among user characteristics, system-induced effects, and miscalibration in existing recommender systems}: We investigate the presence of system-induced effects in existing movie recommendation systems, as well as their associations with user characteristics and miscalibration. Based on the proposed measures, we characterize the behavior of recommender systems that tend to stereotype user preferences and discriminate against atypical users and minority groups. Our analysis reveals three key findings with implications for mitigating this issue. 
    \item \textbf{An investigation into the problem of data underrepresentation}: To explore the possibility of mitigating the disparate impacts of systematic behaviors over minority groups and individuals, we analyze the effect of alleviating data underrepresentation by oversampling users who are stereotyped and explore how it changes the systematic behavior of recommender systems. We discuss its effect on the amelioration of stereotyping and miscalibration, as well as the remaining challenges.
\end{itemize}


\section{Related Work} 
\subsection{Bias and discrimination in recommender systems} 
It has been studied \cite{ControllingPopularityBiasLearningtoRank, ExposureIdeologicallyDiverseNews, Howalgorithmicconfoundingrecommendationsystems, sunstein2009going} that recommender systems are susceptible to biases that favor popular items or users, or get stuck on items and users of the similar kinds. Among them, popularity bias \cite{ControllingPopularityBiasLearningtoRank, hitsnicheshowpopularartists} is a widely studied systematic behavior unfavorable towards less popular items in the long tail appearing much less than popular items. These biases can be amplified while in the iterative feedback loop \cite{Howalgorithmicconfoundingrecommendationsystems}, leading to rich-getting-richer effect. Studies have captured its consequential effects, such as filter bubbles \cite{UsercontrollableRecommendationFilterBubbles, ExposureIdeologicallyDiverseNews, sunstein2009going} where homogeneous items are being over-recommended, or echo chambers \cite{Exploringfilterbubbleeffectusing, sunstein2009going} where groups are more polarized in their preferences or ideologically segregated. These in turn can degrade the quality of recommended items or prevent users from discovering new items.

These biases also tend to have a disparate impact over demographic groups and individuals, discriminating against minor groups or users by offering a lower quality of recommendations. Recent studies have pointed out that these fairness problems in recommender systems are even more complex due to its multi-sided nature \cite{Exploringauthorgenderbookrating} in either or both consumer and provider side. For example, in book recommendations, female authors were found to receive lower exposure to customers than male authors \cite{Exploringauthorgenderbookrating}. This exposure bias was found to be related to the size or entropy of the user rating profile in \cite{InvestigatingPotentialFactorsAssociatedGender}. According to the analysis, females tend to have fewer interactions and specific interests, and as a result, receive lower-quality recommendations. While various biases have been studied, our study attempts to view these system-induced biases from the perspective of underrepresentation problem. Focusing on the user side, we investigate how individuals and groups who are underrepresented in the data have a greater impact of bias and discrimination in recommender systems.

\subsection{Stereotyping problem in recommendations}
Stereotype is a widely known cognitive phenomenon that refers to people’s shared beliefs over specific group characteristics. While stereotype can help an intuitive assessment of groups \cite{StereotypesStereotypingMoralAnalysis}, it can easily lead to overgeneralization of individuals' characteristics \cite{Fairnessrepresentationquantifyingstereotypingrepresentationala} based on their group membership. For example, in the movie preferences,   \cite{TearsFearsComparingGenderStereotypes} found that people tend to cognitively exaggerate group differences with respect to widely known gender stereotypes (e.g., men like action and sci-fi movies) greater than how groups actually differ. 
Recent research has explored the possibility that machine learning or recommender systems may also be subject to stereotypes. On the bright side, previous research has attempted to utilize stereotype-based user modeling in several different recommender system domains \cite{GenderstereotypereinforcementMeasuringgender, UnequalRepresentationGenderStereotypesImage, stereotypemostpopular, evaluatingstereotypenonstereotype}. On the other hand, some of the existing literature \cite{improvingnewuser, evaluatingstereotypenonstereotype} argued that a trained classifier can pick up biases that warp the representational space by pulling outlying examples closer to the prototypical preferences of a dominant group. Recent research \cite{Crankvolumepreferencebiasamplification, BiasDisparityRecommendationSystems, BiasDisparityCollaborativeRecommendationAlgorithmic} on recommender systems has investigated bias disparity, a problem related to the stereotyping issue at the category level.  It refers to a disparity or amplification of users' preference ratio on a particular item category. In \cite{Crankvolumepreferencebiasamplification, BiasDisparityRecommendationSystems, BiasDisparityCollaborativeRecommendationAlgorithmic}, collaborative filtering algorithms were examined to observe bias disparity especially in the exclusively favored genres by males or females such as action and romance movies. To mitigate these biases, several approaches have been proposed to construct and intervene in the causal pathways of user preferences \cite{DeconfoundedRecommendationAlleviatingBiasAmplification} or train recommender systems with dual representations of user preferences \cite{MinorityMattersDiversityPromotingCollaborativeMetric, StereotypingProblemCollaborativelyFilteredRecommender}. While these studies shed light on the existence of stereotypes, the stereotyping problem in recommender systems has not been thoroughly defined and examined in terms of its effect and relationship with user characteristics. In this study, we introduce the notion of system-induced stereotype and investigate how it affects underrepresented groups and individuals.

\subsection{Miscalibration}
Miscalibration in recommender systems refers to the discrepancy between actual and predicted preferences over item categories \cite{Calibratedrecommendations}. Compared to item-level metrics, such as Normalized Discounted Cumulative Gain (nDCG), measuring the accuracy of recommendations in top-ranked items, this category-level measure allows evaluating a system's capability of distributional calibration across various areas of interest (i.e., item categories). In the later work, miscalibration was investigated in terms of its relationship with bias and fairness \cite{AllCoolKidsHowThey, ConnectionPopularityBiasCalibrationFairness, ImpactPopularityBiasFairnessCalibration}. In \cite{ImpactPopularityBiasFairnessCalibration}, they found that a greater impact of popularity bias can lead to an increase in miscalibration. Such an impact was also found to vary across recommendation algorithms and demographic groups. 

On the other hand, little has investigated the deeper understanding of miscalibrations, i.e., how it is associated with different sources of error and system-induced effects. Our study introduces a framework for decomposing miscalibration into bias, variance, and stereotype, which allows one to disambiguate how different systematic effects are combined to form the overall miscalibration and how to mitigate them.

\subsection{Diversity} 
Diversity is one of the most highly valued properties in recommender systems, as it allows users to be presented with a variety of unexpected and diverse items. Measures such as intra-list distance \cite{Improvingrecommendationliststopicdiversification}, or entropy \cite{LearningRecommendAccurateDiverseItems, ChallengingLongTailRecommendation}, have been extensively used in studying diversity and its bias effect in recommender systems. Several studies have investigated how diversity can be promoted with shared accessibility of user preference representations \cite{MinorityMattersDiversityPromotingCollaborativeMetric} or by iteratively updating learning objective for diversity in a supervised \cite{LearningRecommendAccurateDiverseItems} or unsupervised manner \cite{ChallengingLongTailRecommendation}. In our study, we observe that increasing diversity in a recommender system can result in unexpectedly poor recommendations; we refer to this phenomenon as ``inflated diversity.'' We find that diverse items that are liked by typical users are recommended to atypical users with specific interests. Throughout the study, we investigate how this system-induced effect is related to other measures and how it disproportionately affects groups and individuals.

\section{Miscalibration, bias, and stereotype}
\label{sec:method}

\begin{figure}
    \includegraphics[width=\columnwidth]{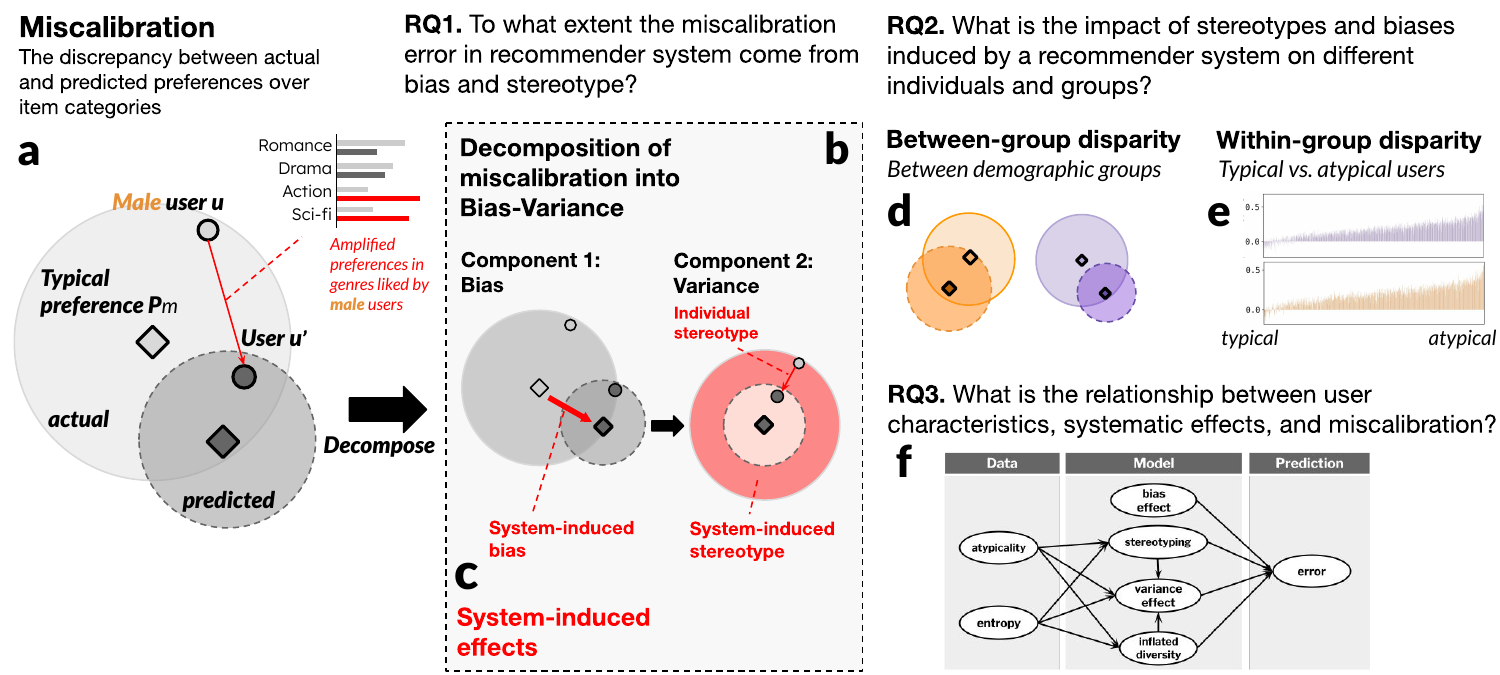}
    \caption{\label{fig:intro-overview}
    The overview of the study. We aim to provide a unified framework for the systematically examining  miscalibration, system-induced bias and stereotype. (a) Miscalibration is a category-level error in recommender systems, referring to the discrepancy in distributions between the distributions of actual and predicted preferences over item categories. (b) Our framework proposes to decompose miscalibration into two distinct sources of error, bias and variance. (c) It allows us to capture how two system-induced effects, bias and stereotype, can be associated with miscalibration. (d, e) By measuring the group-level or individual-level disparity, we measure whether miscalibration, bias, and stereotype disproportionately impact groups and individuals. (f) The relationship between user characteristics, system-induced effects, and miscalibration can be better understood through association analysis.
    }
\end{figure}

In this section, we present a unified framework for systematic understanding of miscalibration, bias, and stereotype in recommender systems. Fig. \ref{fig:intro-overview}) provides an overview of our research methodology. As illustrated in Fig. \ref{fig:intro-overview}a, the process of generating recommendations can be conceptualized as the mapping between the space of users' actual preferences (colored as light gray) and the space of users' predicted preference (colored as dark gray). A miscalibration problem arises when a recommender system fails to predict a user's preference of items at the category level, e.g., a user who does not favor sci-fi movies frequently receives sci-fi movie recommendations. Conceptually, such a mismatch can be understood as a misalignment between two spaces in Fig. \ref{fig:intro-overview}a. In our framework, this distributional misalignment can be decomposed into two distinct parts of distortions, bias and variance (Fig. \ref{fig:intro-overview}b). Inspired by the well-known Bias-Variance decomposition in machine learning literature \cite{BiasVarianceDecompositionBayesianDeepLearninga, biasvariancegeneralloss}, we develop a statistical framework that decomposes miscalibration into bias and variance and further characterizes the system-induced stereotype effect. We show that miscalibration is associated with system-induced bias and stereotype (Fig. \ref{fig:intro-overview}c). These effects can disproportionately affect groups (Fig. \ref{fig:intro-overview}d) and individuals (Fig. \ref{fig:intro-overview}e), resulting in discrimination in terms of recommendation quality. 

In the following, we propose a set of measures to allow a deeper understanding of miscalibration. In Section \ref{sec:miscalibration}, we first define the notion of miscalibration error. Next, we show that the miscalibration can be decomposed into its two sources of error using Bias-Variance decomposition under KL-divergence loss (Section \ref{sec:bias-variance-decomposition}). This allows us to discuss system-induced bias and stereotype as two distinct system-induced effects that are related to bias and variance respectively (Section \ref{sec:system-induced-effects}). 

\subsection{Miscalibration}
\label{sec:miscalibration}
As first introduced in \cite{Calibratedrecommendations}, miscalibration quantifies the discrepancy in distributions between an individual's actual preference $p(\vec{c}|u)$ and predicted preferences $q(\vec{c}|u)$ over item categories $\vec{c}$ (e.g., movie genres) for a user $u$:

\begin{itemize}
    \item $p(\vec{c}|u)$: The distribution over categories $\vec{c}$ of the set of items interacted by a user (i.e., actual preference).
    \item $q(\vec{c}|u)$: The distribution over categories $\vec{c}$ of the set of recommended items in a user (i.e., predicted preference).
\end{itemize}

The individual-level miscalibration, $MC_u$, for a user $u$, uses Kullback-Lieber divergence $D_{KL}$ to calculate the discrepancy between two probability distributions to measure how $q$ deviates from $p$ \cite{Calibratedrecommendations}, as follows: 

\begin{equation}
    MC_u(p, q) = D_{KL}(p||q) = \sum_{c}{p(c|u)} \log \frac{p(c|u)}{q(c|u)}
    \label{eq:miscalibration}
\end{equation}

Miscalibration in the range of $[0, \inf)$ gets smaller when $q$ and $p$ are similar. The zero miscalibration indicates perfect calibration. Similarly, the system-level miscalibration can be denoted as $MC(P, Q)$, where $P$ and $Q$ indicate the sets of actual and predicted preferences for all users.

\begin{table}[]
\scalebox{0.9}{
\begin{tabular}{llll}
\toprule
\textbf{Notation} &
  \textbf{Term} &
  \textbf{Definition} &
  \textbf{Type} \\ \midrule
  \begin{tabular}[c]{@{}l@{}}$p$, $q$ \\ \newline \\ \newline \end{tabular} &
  \begin{tabular}[c]{@{}l@{}}Actual/predicted \\ preference \\for a user $u$ \end{tabular} &
  \begin{tabular}[c]{@{}l@{}}Probabilistic distribution of a user $u$'s \\ actual/predicted preference over item categories $\vec{c}$ \\ \newline \end{tabular} &
  \multirow[t]{2}{*}{\begin{tabular}[c]{@{}l@{}}User\\ preference \\ \newline \end{tabular}}  \\
  \begin{tabular}[c]{@{}l@{}}$P$, $Q$ \\ \newline \\ \newline \end{tabular} &
  \begin{tabular}[c]{@{}l@{}}Actual/predicted \\ preferences \\ for users $U$ \end{tabular} &
  \begin{tabular}[c]{@{}l@{}}Probabilistic distribution of users $U$'s \\ actual/predicted preference over item categories $\vec{c}$ \\ \newline \end{tabular} & \\
\begin{tabular}[c]{@{}l@{}}$\bar{P}$, $\bar{Q}$ \\ \newline \\ \newline \end{tabular} &
  \begin{tabular}[c]{@{}l@{}}Typical \\ actual/predicted \\ preferences\end{tabular} &
  \begin{tabular}[c]{@{}l@{}}Mean actual/predicted preferences $P$, $Q$ of users $U$ \\ over categories $\vec{c}$ \\ \newline \end{tabular} &
   \\ \midrule

$AT_u$ &
  Atypicality &
  Deviation of a user’s preference from typical preference &
  \multirow{2}{*}{\begin{tabular}[c]{@{}l@{}}User\\ characteristics\end{tabular}} \\
$DV_u$ &
  Diversity &
  Variance of a user preference over item categories $\vec{c}$ &
   \\ \midrule

\begin{tabular}[c]{@{}l@{}}$MC_u(p, q)$ \\ \newline \end{tabular} &
  \begin{tabular}[c]{@{}l@{}}Individual-level \\ miscalibration \end{tabular} &
  \begin{tabular}[c]{@{}l@{}}Individual-level discrepancy between a user $u$'s actual \\ preference $p$ and predicted preference $q$ \end{tabular} &
  \multirow[t]{2}{*}{\begin{tabular}[c]{@{}l@{}}System\\ error \end{tabular}} \\
\begin{tabular}[c]{@{}l@{}}$MC(P, Q)$ \\ \newline \end{tabular} &
  \begin{tabular}[c]{@{}l@{}}System-level \\ miscalibration\end{tabular} &
  \begin{tabular}[c]{@{}l@{}}System-level discrepancy between actual preferences $P$ \\ and predicted preferences $Q$ over users $U$  \end{tabular}  &
   \\ \midrule

\begin{tabular}[c]{@{}l@{}}$BE_u(p, q)$ \\ \newline \\ \newline \end{tabular}  &
  \begin{tabular}[c]{@{}l@{}}Individual-level \\ bias effect \\ \newline \end{tabular}  &
  \begin{tabular}[c]{@{}l@{}}The amount of changes in prediction error for a user \\ $u$ when using the predicted mean preference $\bar{Q}$ \\ instead of actual mean preference $\bar{P}$\end{tabular} & 
  \multirow[t]{2}{*}{\begin{tabular}[c]{@{}l@{}}System \\-induced\\ effect \end{tabular}} \\
\begin{tabular}[c]{@{}l@{}}$B(P, Q)$ \\ \newline \end{tabular} &
  \begin{tabular}[c]{@{}l@{}}System-level bias \\ \newline \end{tabular} &
  \begin{tabular}[c]{@{}l@{}}Systematic distortion in distributions \\ between typical actual/predicted preferences $\bar{P}$\end{tabular} &
   \\
   \begin{tabular}[c]{@{}l@{}}$ST$ \\ \newline \end{tabular} &
  \begin{tabular}[c]{@{}l@{}}System-level \\ stereotype\end{tabular} &
  \begin{tabular}[c]{@{}l@{}}Systematic overgeneralization of predicted \\ preferences $Q$ over users $U$ \end{tabular} &
   \\
 \begin{tabular}[c]{@{}l@{}}$ST_u$ \\ \newline \end{tabular} &
  \begin{tabular}[c]{@{}l@{}}Individual-level \\ stereotype\end{tabular} &
  \begin{tabular}[c]{@{}l@{}}Systematic overgeneralization of a predicted  \\preference $q$ for a user $u$ \end{tabular}  &
   \\
   \begin{tabular}[c]{@{}l@{}}$IDV_u$ \\ \newline \end{tabular} &
  \begin{tabular}[c]{@{}l@{}}Individual-level\\ inflated diversity  \end{tabular} &
  \begin{tabular}[c]{@{}l@{}}Systematic inflation of a user's predicted \\preference $q$ for a user $u$ \end{tabular} &
   \\ \bottomrule
\end{tabular}
}
\label{tbl:notation}
\caption{\label{tbl:notation} Mathematical notion, term, definition, and type of concepts in this study.}
\vspace{-1em}
\end{table}

\subsection{Bias-Variance Decomposition for miscalibration}
\label{sec:bias-variance-decomposition}
In this section, we present the statistical measure for decomposing miscalibration into bias and variance. We employ the mathematical equation of Bias-Variance decomposition under Kullback-Lieber divergence loss \cite{biasvariancegeneralloss, biasvairancebayesiandeeplearning}, where each user's preferences are represented as a probabilistic distribution over multiple categories of items, and the loss indicates the discrepancy between the distributions of the user's actual and predicted preferences. 

\textbf{System-level bias and variance.} First, we show that the miscalibration can be decomposed into bias and variance for all users, which follows the decomposition of Kullback-Leibler divergence as introduced in \cite{BiasVarianceDecompositionBayesianDeepLearninga, biasvariancegeneralloss}:

\begin{equation}
    MC(P,Q) = \underbrace{D_{KL}(\bar{P}||\bar{Q})}_\text{Bias} + \underbrace{\mathds{E} [D_{KL}(\bar{Q}||Q)}_\text{Variance}].
    \label{eq:BV}
\end{equation}

In Eq. \ref{eq:BV}, two distributions $P$ and $Q$, which refer to as the data and predictions in the general machine learning setting, can be translated as the set of actual/predicted preferences over users $U$. In this context, two notations $\bar{P}$ and $\bar{Q}$ indicating the mean of the distributions $P$ and $Q$ can be referred to as the \textit{typical} preference. Miscalibration can be decomposed into two sources of error: (i) {\bf Bias}: the discrepancy between the means of actual preferences and predicted preferences (i.e., the discrepancy between typical preferences). It quantifies the part of error \textit{between} actual and prediction in terms of typical preference. (ii) {\bf Variance}: the total variability of predicted preferences. This quantifies another part of the error \textit{within} predicted preferences, which is how much individual users' predicted preferences deviate from the the typical preference (i.e., mean preference).

\textbf{Individual-level bias and variance effect.} Next, we introduce statistical measures of bias and variance effect at the individual level. Different from the system-level bias and stereotype, this measure quantifies how miscalibration for each user is associated with bias and variance at the individual level. Inspired by the decomposition of general loss into bias-variance effect \cite{BiasVarianceDecompositionsLikelihoodBasedEstimators, biasvariancegeneralloss}, we propose individual-level bias and variance effects under KL-divergence loss as follows: 

\begin{equation}
    MC_u(p,q) = \underbrace{(D_{KL}(p||\bar{Q}) - D_{KL}(p||\bar{P}))}_\text{Bias effect}
    + \underbrace{(D_{KL}(p||q) - D_{KL}(p||\bar{Q}))}_\text{Variance effect},
    \label{eq:BV-effect}
\end{equation}

where $p$, $q$ indicate the actual and predicted preferences for an individual user $u$. Following the definitions in \cite{biasvariancegeneralloss}, individual-level miscalibration can be decomposed into  two terms defined as follows : (i) {\bf Individual bias effect}: the amount of changes in prediction error for a user $u$ when using the predicted mean preference (i.e., $\bar{Q}$) instead of the actual mean preference (i.e., $\bar{P}$). (ii) {\bf Individual variance effect}: the discrepancy between prediction errors with and without variability in a given prediction. More specifically, a prediction $q$ can differ across users, whereas there is no variability if all users receive the same prediction according to the mean $\bar{Q}$.

\textbf{Distinction between system-level effect (Eq. \ref{eq:BV}) and individual-level effect (Eq. \ref{eq:BV-effect}).} In contrast to the miscalculation decomposed at the collective level (Eq. \ref{eq:BV}), Eq. \ref{eq:BV-effect} specifically looks into the bias and variance decomposition of the individual-level miscalculation by introducing a user's actual and predicted preferences, namely $p$ and $q$. The specification follows a difference-in-differences paradigm: The individual bias effect measures the difference in the differences between a user's actual preference and the mean actual preferences and between the user's actual preference and the mean predicted preferences. The individual variance effect, on the other hand, measures the difference in the differences between a user's actual and predicted preferences, and between the user's actual preference and the mean predicted preferences. Both can take a positive or negative value, which contributes to the increasing or decreasing miscalibration for a given user.

\subsection{System-induced effects}
\label{sec:system-induced-effects}
Compared to Section \ref{sec:bias-variance-decomposition}, the measures on system-induced effects presented in this section highlight two types of quantities of errors relating to bias and variance respectively.

\textbf{System-level bias and stereotype.} At aggregated level, two system-level effects quantify the overall errors introduced by the model across all users. 

\begin{itemize}
    \item \textbf{System-level bias}: Systematic distortion between mean actual and predicted preferences. This is equivalent to overall bias in Eq. \ref{eq:BV} that can be denoted as $B(P,Q)= D_{KL}(\bar{P}||\bar{Q})$.
    
    \item \textbf{System-level stereotype}: Systematic overgeneralization of predicted preferences based on certain characteristics of individuals, defined as follows:

    \begin{equation}
        ST(P,Q) = 1 -  \frac{\mathds{E}_{q \in Q} [D_{JS}(q||\bar{Q})]}{\mathds{E}_{p \in P} [D_{JS}(p||\bar{P})]},
    \end{equation}

    where the mean deviations of actual and predicted preferences $P$ and $Q$ for users $U$ from typical preference $\bar{P}, \bar{Q}$ are computed as $\mathds{E}_{p \in P} [D_{JS}(P||\bar{P})$] and $\mathds{E}_{q \in Q} [D_{JS}(Q||\bar{Q})]$ respectively. We use Jensen-Shannon divergence to compute the deviation for individual user's preference $p$ from typical preference $\bar{P}$, where $D_{JS}(p||\bar{P})=\frac{D_{KL}(p||\bar{P})+D_{KL}(\bar{P}||p)}{2}$ is a symmetric measure of quantifying the distance between two probabilistic distributions. The system-level stereotype measures the relative difference between two quantities, i.e., how the variance in actual preferences decreased in predicted preferences, which has a range of $(-\inf, 1)$. When $ST(P,Q) > 0$, it indicates user preferences in recommendations stereotyped towards typical preference at the system-level. The zero quantity in $ST(P,Q)$ indicates no overall stereotypes.
\end{itemize}

\textbf{Individual-level stereotype.} The stereotype at the individual level is defined as follows:

\begin{equation}
    ST_u(p,q) = D_{JS}(p||\bar{P}) - D_{JS}(q||\bar{Q}).
    \label{eq:indi-st}
\end{equation}

\textbf{Individual-level inflated diversity.} We introduce the concept of ''inflated diversity'' to describe the situation in which a user receives an over-diverse prediction $q$, i.e., one that is more diverse than the user's actual preference $p$. It quantifies how a user with particular interests is forced to resemble dominant groups with diverse preferences.    

\begin{equation}
    IDV_u(p, q) = DV_u(p) - DV_u(q),
\end{equation}

where $DV_u$ measures the diversity of a user $u$ (see Eq. \ref{eq:diversity}).

\textbf{Bias disparity.} Introduced by \cite{BiasDisparityRecommendationSystems}, bias disparity measures the relative difference in the preference ratio of users $U$ on an item category $c$ between actual and predicted preference.

\begin{equation}
    BD(c, U) = 1 - \frac{q(c, U)}{p(c, U)},
\end{equation}

where $p(c, U)$ and $q(c, U)$ denote the distribution of actual and predicted preferences over an item category $c$ of the set of items for users U. When $BD(c, U) > 0$, it indicates the amplification of preference ratio on item category $c$ in predicted preference compared with actual preference. We leverage this measure to interpret system-induced effects at category level.

\subsection{User characteristics}
\label{sec:user-characteristics}
We further consider two types of user characteristics, atypicality and diversity, which are related to the variability of user preferences over item categories. 

\textbf{Atypicality.} This quantifies the deviation of a user's preference from typical preference, which can be measured by the distance between two distributions, typical preference and a user's preference:

\begin{equation}
    AT_u = D_{JS}(p||\bar{P}).
\end{equation}

Similarly, the predictive atypicality in the range of $[0, 1]$ can be defined by replacing the observed quantities $p$ and $\bar{P}$ with the two predicted quantities $q$ and $\bar{Q}$, respectively.
 
\textbf{Diversity.} This measures the variance of individual preferences:

\begin{equation}
    DV_u = Entropy(u) = -\sum_{c \in \vec{c}} p(\vec{c}|u) \log p(\vec{c}|u),
    \label{eq:diversity}
\end{equation}

where $Entropy$ denotes the normalized entropy of preference distribution for an individual user $u$, which has a range of $[0, 1]$. A higher diversity indicates a user preference varying over multiple categories to a similar degree. A lower diversity indicates that a user prefers certain categories particularly more than others.
\section{Analysis results}
Based on the proposed measurement framework, we investigate the miscalibration, system-induced effects, and disparate impact of recommender systems. We examine algorithmic, group, and individual differences in the context of movie recommendation in order to answer the following three analytic questions: 

\begin{itemize}
    \item \textbf{RQ1.} To what extent do two distinct sources of error, bias, and variance, contribute to miscalibration?
    \item \textbf{RQ2.} What is the impact of stereotypes and biases induced by a recommender system on different individuals and groups?
    \item \textbf{RQ3.} What is the relationship between user characteristics, systematic effects, and miscalibration?
\end{itemize}

\begin{figure}
    \includegraphics[width=.9\textwidth]{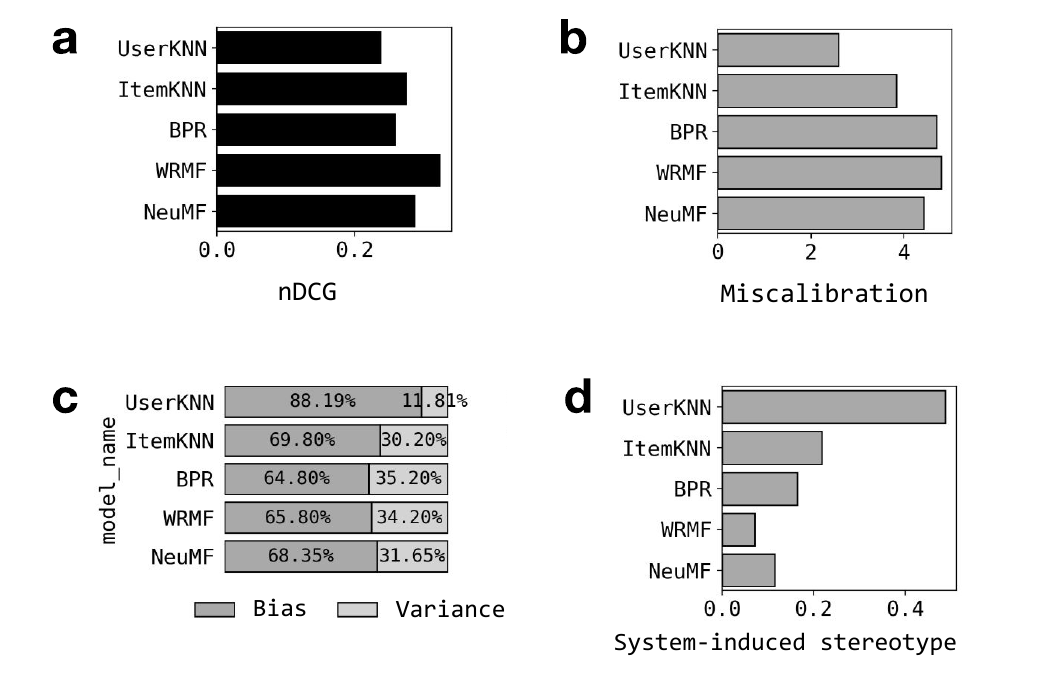}
    \caption{\label{fig:result-algorithm}
    The algorithmic difference of system-induced effects and recommendation quality over five recommendation algorithms. Two performance results in (a) nDCG@20 and (b) miscalibration@20 show that complex models tend to perform better for item-wise predictions (nDCG@20) but worse at category-wise calibration (miscalibration@20). (c) In Bias-Variance ratio, complex models tend to have a higher variance, indicating the capability of better approximating the variance in actual preference. (d) This in turn leads to lower system-induced stereotype.
    }
\end{figure}

\subsection{Experimental setting}
\label{sec:experimental-setting}

\textbf{Data.} In the analysis, we use MovieLens 1M dataset\footnote{https://grouplens.org/datasets/movielens/1m/} that contains demographic attributes such as gender and age. The original dataset includes 1,000,209 user ratings ranging from 1 to 5 over movie items across 18 movie genres. We take three preprocessed steps frequently adopted in existing work \cite{Crankvolumepreferencebiasamplification, DeconfoundedRecommendationAlleviatingBiasAmplification, ConnectionPopularityBiasCalibrationFairness} to follow the best practice and ensure our work comparable to them: (1) We binarized ratings by dropping all user-item records with ratings less than 4 to convert it to implicit feedback. Then, (2) we remove all users with less than 20 interactions to ensure the quality of preference information of each individual. Lastly, (3) for items associated with more than multiple genres, we normalized scores in each item to be proportional to the number of per-category multiple categories. After the preprocessing steps, the data includes 562,800 interactions from 5,180 users and 3,526 items.

Following the experiment settings in previous literature \cite{ConnectionPopularityBiasCalibrationFairness, Calibratedrecommendations}, the implicit feedback data was split into 80\% and 20\% for training and testing using user-fixed setting in Librec-auto package where the interactions within each user were divided into training and test set with respect to the ratio based on chronological order. Experiments for all five algorithms (details below) were conducted with the same train-test split data. We employ the same setting to facilitate the comparison of all results derived from predicted recommendations.

\textbf{Algorithms.} In the experiment, we compare five recommendation algorithms to examine the generalizability of our observation on system-induced effects. Our choice of algorithms, based on a survey on related work in line of research, is to ensure a variety of recommendation algorithms to be examined, encompassing four types of methods ranging from the baselines to state-of-art algorithms, including memory-based (ItemKNN \cite{itemknn}, UserKNN \cite{userknn}), learning-to-rank (Bayesian Personalized Ranking, BPR \cite{bpr}), latent factor (Weighted Regularized Matrix Factorization, WRMF \cite{wrmf}) and deep learning model (Neural Matrix Factorization, NeuMF \cite{neumf}). These five recommender systems were experimented with hyperparameters with the highest performance based on the hyperparameter search supported in the Librec-auto package (UserKNN, ItemKNN, BPR, and WRMF using Librec-auto\footnote{https://github.com/that-recsys-lab/librec-auto} \cite{mansoury2018automating, mansoury2019algorithm} and NeuMF using QRec\footnote{https://github.com/Coder-Yu/QRec} for deep-learning model \cite{qrec}, which are python-based framework for recommender systems). The algorithms were evaluated with Normalized Discounted Cumulative Gain (nDCG) \cite{jarvelin2002cumulated} and miscalibration (Eq. \ref{eq:miscalibration}) for top 20 recommended items.

\subsection{RQ1. Algorithmic differences in miscalibration and system-induced effects}
We first analyze the algorithmic differences in miscalibration, system-induced bias, and stereotype, with the goal of understanding how complex and simple models behave differently. Among the five algorithms tested, WRMF and NeuMF are considered more complex models than the others. Fig. \ref{fig:result-algorithm} summarizes the results of algorithmic differences, including: the recommendation performances at the item level (in terms of nDCG, Fig. \ref{fig:result-algorithm}a) and the item-categorical level (miscalibration, Fig. \ref{fig:result-algorithm}b), Bias-Variance ratio as sources of miscalibration (Fig. \ref{fig:result-algorithm}c), and the system-induced stereotype (Fig. \ref{fig:result-algorithm}d). 

The results indicate that complex models, such as WRMF or NeuMF, tend to perform better for item-level predictions but less well for category-level calibration of user preferences.
The two models, UserKNN and NeuMF, are two examples of relatively simple (memory-based) and complex (deep learning method) approaches, respectively. We have three key observations. First, the simplest model, UserKNN, had the lowest performance in nDCG but the least amount of miscalibration (Fig. \ref{fig:result-algorithm}a-b). Second, simpler models have a smaller proportion of variance (Fig. \ref{fig:result-algorithm}c), suggesting that the ratio of variance is positively associated with model complexity. In particular, UserKNN had the lowest variance ratio of the five algorithms examined. Lastly, system-induced stereotypes tend to be more prevalent in simpler models (Fig. \ref{fig:result-algorithm}d). UserKNN exhibited the highest degree of stereotyping, whereas others exhibited a lower degree of stereotyping.


These findings are consistent with a well-known bias-variance trade-off \cite{kohavi1996bias} and its association with model complexity, which also applies to recommender systems: Simple recommendation algorithms are more susceptible to bias, whereas complex models have greater variance. Our analysis also reveals that a higher level of predictive variance is positively correlated with a lower level of stereotype. When a model's predictive preferences more closely resemble the variability of users' actual preferences, the impact of stereotyped predictions on users is reduced.

\begin{figure}
    \includegraphics[width=.9\textwidth]{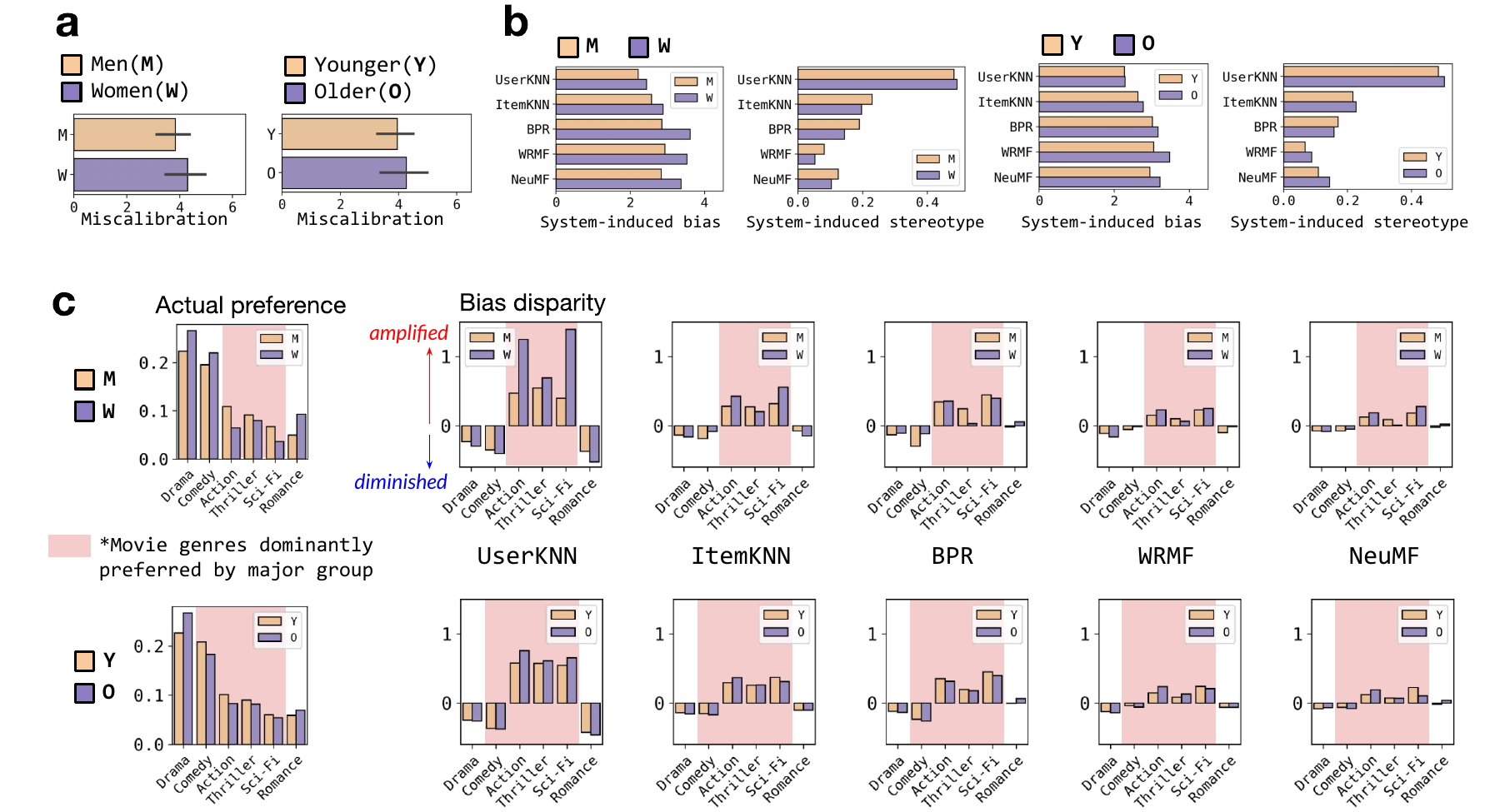}
    \caption{\label{fig:result-grp}
    Disparate impact of miscalibration and system-induced effects over groups.  (a) Miscalibration: For both demographic attributes, gender and age, five recommendation algorithms on average exhibit higher miscalibration for minority groups, women and older users. (b) System-induced effects: Both bias and stereotype tend to impact minority groups to a greater extent. (c) Bias disparity: At the category-level, user preferences in majority-dominant genres (i.e., categories dominantly preferred by majority group) highlighed as red areas tend to be amplified. 
    }
\end{figure}

\subsection{RQ2. Disparate impact of miscalibration, bias, and stereotype over groups and individuals}
We investigate how miscalibration and system-induced effects affect different groups and users disproportionately. We use the statistical measures presented in Section \ref{sec:method} to measure the collective- and individual-level effects in terms of miscalibration, system-induced bias and stereotype.

\subsubsection{Group-level disparity over demographic groups.}
First, we explore the group-level disparity between demographic groups, focusing on two demographic attributes, gender and age (see details for demographic attributes in Section \ref{sec:experimental-setting}).

The disparate impacts of miscalibration and two system-induced effects for both demographic attributes are summarized in Fig. \ref{fig:result-grp}. For miscalibration, two minority groups, women and older users, have a higher degree of miscalibration than majority groups in average over five algorithms (Fig. \ref{fig:result-grp}a), with a significant difference validated in $t$-test ($p \leq 0.001$). Regarding system-induced effects (Fig. \ref{fig:result-grp}b), while system-induced bias showed higher impact on minority groups in both gender and age regardless of algorithms, system-induced stereotype had a mixed result. For the age attribute, older users had a higher level of stereotypes across all algorithms. For gender, on the other hand, women or men users interchangeably had a greater stereotype in different algorithms. For the group disparity at the category level (Fig. \ref{fig:result-grp}c), we observe a consistent pattern of bias disparity (details in Section \ref{sec:system-induced-effects}), which tends to amplify the preference of majority-dominant categories (i.e., categories dominantly preferred by majority group) such as action, thriller, and romance as men-dominant genres and comedy, action, thriller, and sci-fi as genres dominated by younger users (highlighted as red in Fig. \ref{fig:result-grp}c). However, the degree of category-level bias disparity varies across algorithms. Simpler algorithms tend to show higher bias disparity in majority-dominant genres and in minority groups to a greater extent.

The bias disparity tends to increase in majority-dominant categories, such as action, thriller, and romance genres favored predominantly by men, and comedy, action, thriller, and sci-fi genres favored predominantly by younger users. In Fig. \ref{fig:result-grp}c, these categories with a majority dominance are highlighted in red. On the other hand, the degree of the category-level bias disparity varies across algorithms. Simpler models tend to exhibit a greater bias disparity in majority-dominant genres and among minority groups.

\begin{figure}
    \includegraphics[width=\textwidth]{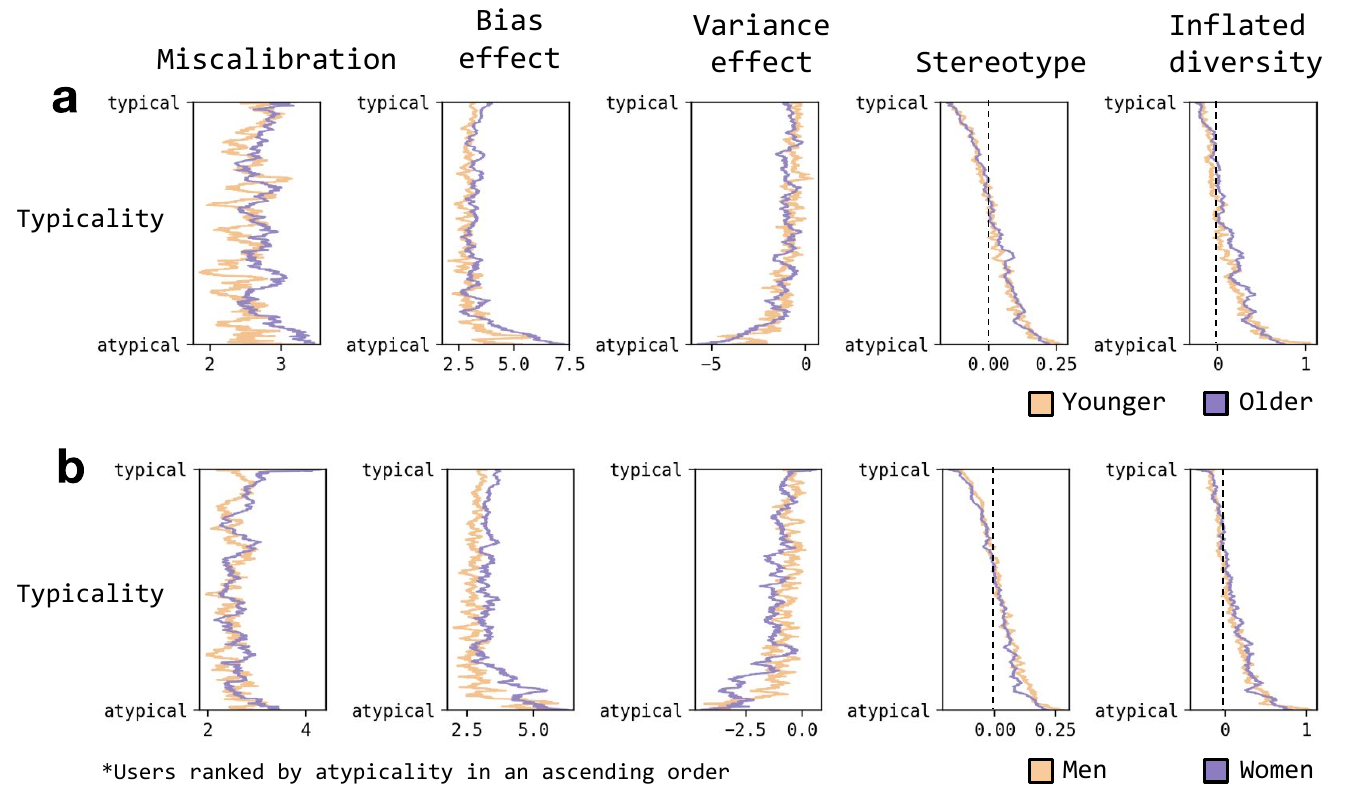}
    \caption{\label{fig:result-individual}
    The individual-level impact of miscalibration, bias/variance effect, stereotype, and inflated diversity in WRMF algorithm for (a) gender and (b) age groups. All individuals are ranked by atypicality in order of the most typical (top) to the most atypical (bottom) users and distinguished by gender (younger and men users in orange; older and women users in purple). 
    }
\end{figure}

\subsubsection{Individual-level disparity between typical vs. atypical users}
Next, we investigate how these system-induced effects impact  typical vs. atypical users disproportionately. This allows us to see how the degree of users' preferences deviated from the average preference can be associated with disparate impacts of system-induced effects and miscalibration.

Fig. \ref{fig:result-individual} shows the individual-level system-induced effects, including (from left to right) miscalibration, bias and variance effects, stereotyping, and inflated diversity, where users are sorted according to their typicality (top: the most typical users, bottom: the most atypical users). The two curves represent the trends of system-induced effects over user typicality for different age (Fig. \ref{fig:result-individual}a) and gender (Fig. \ref{fig:result-individual}b) groups.

We see a clear difference between typical and atypical users in terms of the impact of stereotype and inflated diversity. First, while there is no discernible difference between typical and atypical users in terms of miscalibration, the bias and variance effect impact these user groups differently. In particular, atypical users have a greater bias effect and a lower variance effect than typical users. Second, stereotyping and inflated diversity, two system-induced effects at the individual level, tend to be positively associated with atypicality. Users with more specific preferences at the bottom of the plots, as shown in Fig. \ref{fig:result-individual}, tend to be more stereotyped and have a higher degree of inflated diversity. In contrast, typical users tend to experience the opposite effect, with their preferences becoming inversely stereotyped and less diverse. This indicates that the recommendations received by atypical users tend to be overgeneralized as a result of the combined effects of stereotyping and inflated diversity.

\subsection{RQ3. Relationship between user characteristics, system-induced effects, and miscalibration}

In the previous sections, we discovered that recommendation algorithms consistently have a disparate impact on groups and individuals, even though they exhibit miscalibration and system-induced effects to varying degrees. As a result, minority groups and atypical individual users were discriminated against. Given all of the findings from RQ1 and RQ2, we intend to systematically investigate the relationship between user characteristics, system-induced effects, and miscalibration.

\begin{figure}
    \includegraphics[width=.9\textwidth]{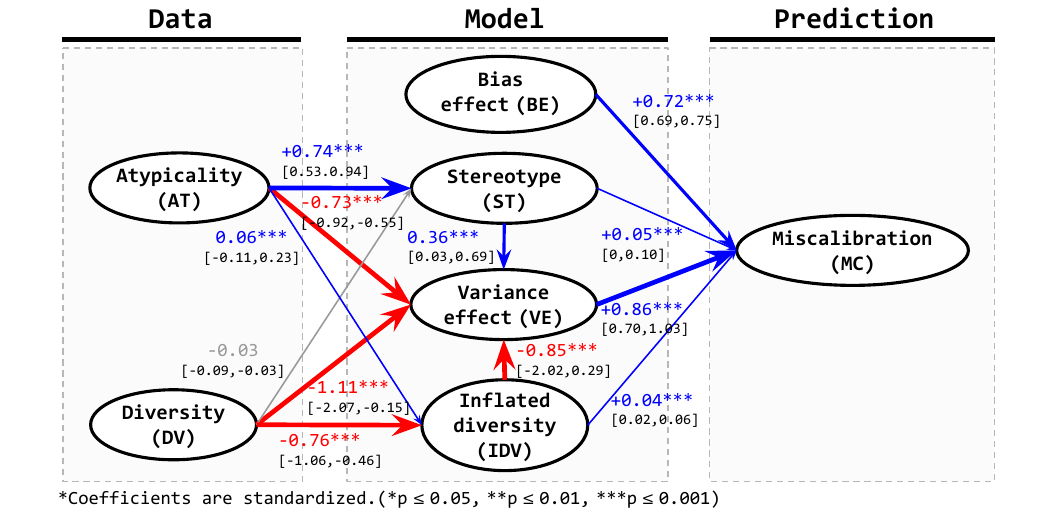}
    \vspace{-1em}
    \caption{\label{fig:result-sem}
    The relationship between systematic effects and miscalibration across five recommendation algorithms.
    }
    \label{fig:sem}
\end{figure}

We hypothesize that a recommender system's miscalibration (between users and their preferred genres) is a combined result of the various forms of bias produced by its model, trained with data comprising different kinds of users. Fig. \ref{fig:sem} illustrates our hypothesis. We consider two user factors that can be directly measured from the input data ({\bf Data} phase), atypicality and diversity, four types of system-induced effects (individual-level bias, variance, stereotype, and inflated diversity) that are produced by a model ({\bf Model} phase), and the prediction performance evaluated in terms of miscalibration ({\bf Prediction} phase). We use Structural Equation Modeling (SEM) \cite{ullman2012structural} to statistically test the associations between different factors, as it allows us to test the direct and indirect relationships among the factors.

\begin{figure}
    \includegraphics[width=0.9\columnwidth]{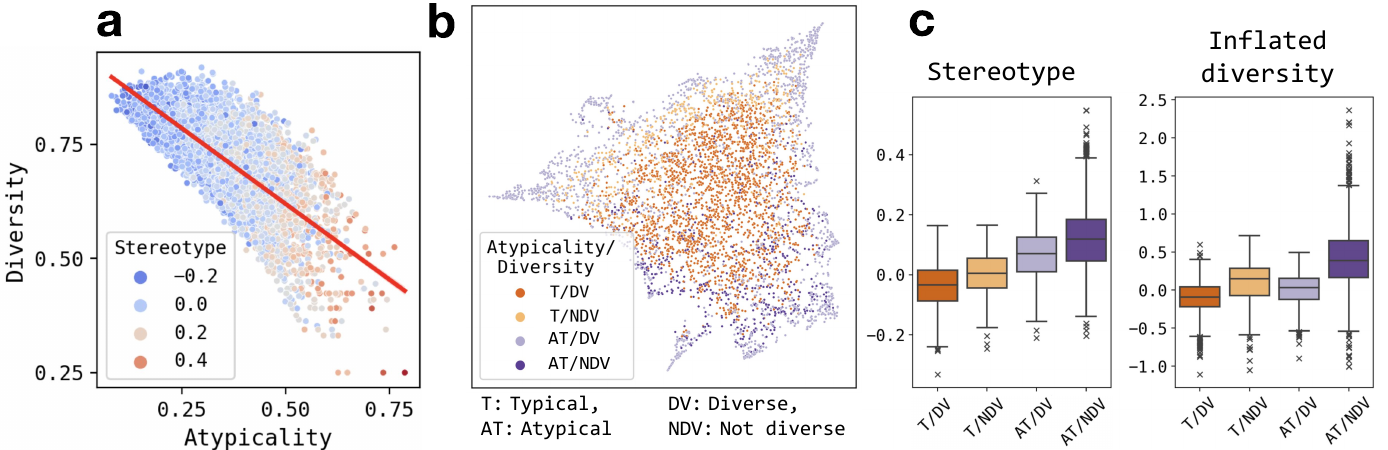}
    \caption{\label{fig:result-sem-user}
    The relationship between systematic effects and miscalibration across five recommendation algorithms. (a) Atypicality and diversity are negatively associated with each other. This indicates that atypical users tend to have particular and less diverse preferences. (b) Four groups of users based on their atypicality and diversity are clearly distinct over user space, confirming the relationship between users' deviation from and two user characteristics. (c) Four groups of users are significantly different in two system-induced effects. More atypical and diversity users are found to have higher impact of stereotype and inflated diversity.  
    }
    \label{fig:user-charcteristics}
\end{figure}

The result of structural equation modeling is summarized in Fig. \ref{fig:result-sem}. All paths are annotated with its standardized coefficient of the mean and 95\% confidence interval from SEMs across five recommendation algorithms. The SEM results revealed several interesting findings.
First, two user characteristics, atypicality and diversity, are negatively associated with each other (Fig. \ref{fig:user-charcteristics}a). This means that typical users tend to prefer diverse movie genres while atypical users tend to have particular interest in certain genres. Second, atypicality and diversity have strong associations with stereotype and inflated diversity respectively: atypicality is positively associated with stereotype indicating atypical users to be more stereotyped; diversity is negatively associated with inflated diversity, meaning that user preferences with a particular interest tend to be forced to be diverse. 

To further look into the relationship between two user characteristics, we divide users in four groups by atypicality (low/high) and diversity (low/high), categorized by thresholding the measures using median value. These four groups of users are visually differentiated in Fig. \ref{fig:user-charcteristics}b, where all users in Movielens 1M dataset were projected into two-dimensional space using UMAP (Uniform Manifold Approximation and Projection) \cite{umap} based on their genre-wise preferences. In the plot, the users with typical and diverse preference (colored as orange) is placed at the center of the user space, while users with atypical and less diverse preference tend to be placed at the periphery. We also find that these four groups exhibit significantly different degree of stereotype and inflated diversity (one-way ANOVA; $p \le 0.001$). As shown in Fig. \ref{fig:user-charcteristics}b, users with atypical and less diverse (AT/NDV) are the most stereotyped and have inflated diversity, while users with typical/diverse are rather inverse-stereotyped and have reduced diversity, indicating the negative association between intersectional user characteristics and two system-induced effects. 

Third, two system-induced effects, stereotype and inflated diversity, were found to have the opposite association with variance effect: higher stereotype tends to increase variance effect (ST $\rightarrow$ VE; 0.36±0.17) while inflated diversity tends to decrease variance effect (IDV $\rightarrow$ VE; -0.85±0.60). From model to prediction phase, variance effect is the most highly associated with miscalibration (VE $\rightarrow$ MC; 0.86±0.10). Lastly, indirect effects from stereotype and inflated diversity to miscalibration also exist. Higher stereotype tends to increase miscalibration (ST $\rightarrow$ VE $\rightarrow$ MC; 0.3±0.1) while higher inflated diversity tend to decrease miscalibration (IDV $\rightarrow$ VE $\rightarrow$ MC; -0.71±0.38), indicating they have an indirect effect on miscalibration.

\section{Mitigating underrepresentation problem}

Throughout R1-R3, we observed that recommendation algorithms tend to exhibit system-induced effects and miscalibration to a different extent but consistently have disparate impacts on minority groups such as women and older users. Bias disparity was also observed to amplify preferences in majority-dominant genres in both gender and age groups. At the individual level, atypical users tended to be stereotyped and have their diversity inflated to a greater extent. All of these findings indicate the systemic effects of overgeneralizing user preferences toward typical/dominant preferences, which impact users who are underrepresented in the data. 

Based on these findings, we examine the possibility of mitigating the disparate impact on the underrepresented users in the data. We hypothesize that modifying the proportion of underrepresented users in the training data will change algorithmic behavior. Thus, our strategy is to use bootstrapped resampling to adjust the data distribution (users' interactions with items) in order to increase the data representation of stereotyped users. We use an equal-sized binning that divides users into $n$ bins as follows. First, we determine the number of bins $n$ where each bin has an equal number of users ($1/n$ of the total number of users). These bins are further split into two groups: the stereotyped ($ST_i > 0$ from Eq. \ref{eq:indi-st}) who will be oversampled and the inverse-stereotyped ($ST_i < 0$ from Eq. \ref{eq:indi-st}) who will not be involved in oversampling into different bins. We determine $n$ such that the inverse-stereotyped users are contained in a single bin, and the remaining bins are the stereotyped users.  

After binning, we determine the oversampling rates $R_k = \{r_k, 1+ b*k\}_{k \in N}$ for stereotyped users, where these rates for $n$ bins are set to linearly increase based on a base sampling rate $b$ between 0.01 to 0.4. Based on oversampling rates, individual users' interaction for each bin is sampled with replacement based on the rate. For oversampling rates greater than 1, we sample only for the portion of a rate $(r_k - 1)$ and then combine the sampled data with the the original data as the portion of 1. After resampling, we train five recommendation models with the updated dataset. We report the results from the original and oversampled training data in Table \ref{tbl:mitigation} that shows the best performance among different base oversampling rates in terms of nDCG@20, Stereotype, and Miscalibration@20.

The results show that oversampling stereotyped users overall helps improve recommendation quality, miscalibration, and stereotype, with its effectiveness over different recommendation models. Specifically, simpler models such as UserKNN or ItemKNN with the highest stereotypes from a larger pool of users have the largest gain of nDCG. On the other hand, NeuMF, the most complex model with the lowest stereotype, has an adverse effect with its decreasing nDCG. We also find that a greater degree of oversampling (i.e., larger base sampling rates) works better for models with greater stereotype.

\begin{table}[]
\begin{tabular}{@{}lllllllll@{}}
\toprule
& \multicolumn{4}{c}{Original} & \multicolumn{4}{c}{Oversampling}                      \\\cmidrule(lr){2-5}\cmidrule(lr){6-9}
 &
  \begin{tabular}[c]{@{}l@{}}\% ST-ed\\ users\end{tabular} &
  \begin{tabular}[c]{@{}l@{}}nDCG\\ @20\end{tabular} &
  \begin{tabular}[tc]{@{}l@{}}ST\\ \end{tabular} &
  \begin{tabular}[c]{@{}l@{}}MC\\ @20\end{tabular} &
  \begin{tabular}[c]{@{}l@{}}nDCG\\ @20\end{tabular} &
  \multicolumn{1}{c}{ST} &
  \begin{tabular}[c]{@{}l@{}}MC\\ @20\end{tabular} &
  \begin{tabular}[c]{@{}l@{}}Base sampling\\ rate\end{tabular} \\ \midrule
UserKNN & 93.5\% & 0.2390          & 0.4708          & \textbf{2.7946} & \textbf{0.2506} & \textbf{0.4667} & 2.8116          & 0.4  \\
ItemKNN & 74.5\% & 0.2751          & \textbf{0.2151} & 3.8714 & \textbf{0.2765} & 0.2373          & \textbf{3.8609}          & 0.15  \\
BPR     & 66.4\% & 0.2598          & 0.1643          & \textbf{4.6925} & \textbf{0.2602} & \textbf{0.1596} & 4.8234          & 0.15 \\
WRMF    & 55.9\% & 0.3251          & 0.0714          & 4.7914          & \textbf{0.3258} & 0.1023          & \textbf{4.7618} & 0.1  \\
NeuMF   & 44.2\% & \textbf{0.2885} & 0.1156          & \textbf{4.4244} & 0.2606          & \textbf{0.0617} & 5.3689          & 0.05 \\ \bottomrule
\end{tabular}
\label{tbl:mitigation}
\caption{\label{tbl:mitigation} The metrics of before (original) and after oversampling stereotyped users.}
\vspace{-2.5em}
\end{table}

While the oversampling technique generally improves stereotyping and miscalibration, its effect varies widely across algorithms and is inconsistent between two measures. When stereotype increases or decreases, miscalibration has the opposite effect for all algorithms. This poses an additional challenge to the effectiveness of mitigating user underrepresentation: system-induced effects and miscalibration have complex interactions in model results.


\section{Limitation and Future Work}

In this section, we discuss the limitations of our work, implications and future work in tackling the problem of system-induced effects in recommender systems. \\

\noindent \textbf{Deeper understanding of user characteristics, system-induced effects, and micalibration.} Our study presents a grounded work to intricately explore and establish connections between various factors in different stages of recommender systems, among user characteristics, system-induced effects and errors. However, we find that the scope of the current study is limited in several ways. First, while our analysis primarily investigated two systematic effects -- stereotype and inflated diversity -- associated with variance effects, we have yet to examine associations related to other factors, such as other system-induced filter bubble or echo chamber effects \cite{ge2020understanding, Exploringfilterbubbleeffectusing}, and their relationship with errors. Those effects are somewhat related to our observations, especially the effect of bias amplification exaggerating preferences over the major categories in our observations, but might be further explained in relation to errors and its components, bias and variance, as examined using our framework. Second, conducting an intersectional analysis across subgroups (e.g., black women) and their category-level preferences may reveal unknown associations, shedding light on the degree to which those subgroups are likely to face system-induced effects, and its relationship with their diversity and atypicality. Lastly, the alignment of system-induced stereotypes with social stereotypes also remains underexplored. Only a few studies have investigated the discrepancy between actual gender difference and social stereotype in some domains such as movie genres \cite{TearsFearsComparingGenderStereotypes}. Such analysis could unveil how data and algorithmic recommendation encode and perpetuate stereotypes minded among people into automatic systems. \\

\noindent \textbf{Mitigation strategies.} To explore the possibility of mitigating system-induced effects, we experimented with an oversampling method -- an approach to overrepresent users who were stereotyped in the dataset. While we used stereotype scores as a threshold for oversampling, the choice of score may depend on how underrepresented users are defined, considering factors like inflated diversity or user characteristics (diversity or typicality).

Despite its overall effectiveness, we observed the mixed effect of our mitigation strategy (i.e., the trade-off between miscalibration and stereotype), indicating the complex relationship between system-induced effects. This could be partially attributed to a challenging nature of mitigating variance effect, indicating that highly varied individual preferences may not be able to be corrected via automated approach. Given this challenge, a user-driven mitigation, allowing users to review miscalibrated and stereotyped items with explanations and interface for intervention, can be a potential direction to address the gap. This approach can further ensure better transparency with a greater degree of user autonomy.

In addition, although our oversampling method effectively mitigates various types of errors and enhance data representation, the approach needs to be used with caution. For instance, it may lead to overfitting towards minority users whose instances are sampled multiple times, resulting in greater model bias, and poor performances on unseen data. \\

\noindent \textbf{Dichotomous analysis.} Our analysis was conducted in binary group settings due to limited data availability (the original dataset including two genders and limited point of ages). A further analysis for multiple groups with available data may lead to different results, which may help generalize our findings to other demographic attributes such as other gender groups or other demographic groups such as ethnicity. \\

\noindent \textbf{Application scope.} Also, future work in different recommendation contexts and more advanced algorithms can also help make our findings more generalizable to a wide range of recommender systems. This may include a diverse type of recommendation ranging from products, online news, and music to insurance and job application.
\section{Conclusion}

In this study, we proposed a unified framework for examining stereotype, bias, and miscalibration. Our framework demonstrated that the problem of miscalibration discussed in previous studies \cite{Calibratedrecommendations, AllCoolKidsHowThey, ConnectionPopularityBiasCalibrationFairness, ImpactPopularityBiasFairnessCalibration} can be better understood by Bias-Variance decomposition, which can then be examined further through two system-induced effects, bias and stereotype, respectively. In the analysis of movie recommendation using the framework, we showed that  recommendation algorithms exhibit systematic behaviors differently in bias and variance, but consistently introduce a higher degree of system-induced bias and stereotype in minority groups and atypical users, and amplify the preference ratio of majority-dominant genres. All of these observations illustrate the systematic tendency for underrepresented users' preferences to be overgeneralized in favor of typical preferences. Our mitigation strategy, the oversampling of stereotypical users, was found to not only enhance recommendation quality but also mitigate the system-induced stereotypes.

\begin{acks}
\end{acks}

\bibliographystyle{ACM-Reference-Format}
\bibliography{references, references-zotero}

\appendix

\end{document}